# Projected Changes in the Iberian Peninsula drought characteristics

García-Valdecasas Ojeda, M., Gámiz-Fortis S.R., Romero-Jiménez, E., Rosa-Cánovas, J.J.,

Yeste, P., Castro-Díez, Y., and Esteban-Parra, M.J.

**ABSTRACT**

High spatial resolution drought projections for the Iberian Peninsula (IP) have been examined in terms of duration, frequency, and severity of drought events. For this end, a set of regional climate simulations was completed using the Weather Research and Forecasting (WRF) model driven by two global climate models (GCMs), the CCSM4 and the MPI-ESM-LR, for a near (2021-2050) and a far (2071-2100) future, and under two representative concentration pathway (RCP) scenarios (RCP4.5 and RCP8.5). Projected changes for these simulations were analyzed using two drought indices, the Standardized Precipitation Evapotranspiration Index (SPEI) and the Standardized Precipitation Index (SPI), considering different timescales (3- and 12-months). The results showed that the IP is very likely to undergo longer and more severe drought events. Substantial changes in drought parameters (i.e., frequency, duration, and severity) were projected by both indices and at both time scales in most of the IP. These changes are particularly strong by the end of the century under RCP8.5. Meanwhile, the intensification of drought conditions is expected to be more moderate for the near future. However, the results also indicated key differences between indices. Projected drought conditions by using the SPEI showed more severe increases in drought events than those from SPI by the end of the century and, especially, for the high-emission scenario. The most extreme conditions were projected in terms of the duration of the events. Specifically, results from the 12-month SPEI analysis suggested a significant risk of megadrought events (drought events longer than 15 years) in many areas of IP by the end of the century under RCP8.5.

# 1. Introduction

The drought phenomenon, characterized mainly by being a period with scarce precipitation, is one of the most devastating natural hazards related to climate change (Kirono et al., 2011; Sheffield and Wood, 2008) with effects in many sectors and systems, such as agriculture, water resources, and natural ecosystems. For southern Europe, there is a recognized consensus about increasing drought conditions during the last decades (Briffa et al., 2009; García-Valdecasas Ojeda et al., 2017; Gudmundsson and Seneviratne, 2015; Spinoni et al., 2015a, 2015b; Vicente-Serrano et al., 2014), being the Mediterranean area considered as an especially vulnerable region (Christensen et al., 2007; Lindner et al., 2010).

In this framework, the Iberian Peninsula (IP), with a highly variable rainfall regime, has presented recurrent droughts and a significant tendency towards more arid conditions in the last decades (Páscoa et al., 2017) fundamentally resulted from an increase in evapotranspiration (Vicente-Serrano et al., 2014). However, in terms of drought trends, the different behaviors found along the 1901-2012 period (Páscoa et al., 2017) highlighted the need to perform analysis at regional scale for the IP (Ficklin et al., 2015). Therefore, the study of drought events requires the use of regional climate models (RCMs) that, driven by global circulation models (GCMs) capture the different processes related to drought episodes at a finer scale. Nevertheless, some difficulties are presented in the RCM simulations associated to the uncertainty caused by different aspects such as the internal variability of the regional model, the parameterization schemes for the model configuration, or errors inherited from the initial and boundary conditions (Lee et al., 2016; PaiMazumder and Done, 2014). As result, the skill of the models in representing historical drought for the IP is very varied (Guerreiro et al., 2017), and the uncertainty introduced by the different simulations should be taken into consideration, especially over those regions that are characterized by a disagreement on the sign of the drought tendency between simulations (Spinoni et al., 2018). Therefore, prior to the use of an RCM for climate change studies, the evaluation of the ability of the RCM to capture regional climate characteristics must be evaluated (Ruiz-Ramos et al., 2016). In this context, different studies have already proved the ability of the Weather Research and Forecasting (WRF) model to capture the behavior of important variables related to drought events in the IP, such as the rainfall (Argüeso et al., 2012a) and the evapotranspiration (García-Valdecasas Ojeda et al., 2020a). Furthermore, the WRF model provides added value in simulating drought conditions over the IP through different drought indices (García-Valdecasas Ojeda et al., 2017).

However, while drought phenomenon over the last decades has been thoroughly studied in the IP, the potential change in future drought remains an element of debate largely attributed to uncertainties related to climate projections (Dai, 2011; Sheffield et al., 2012). For the IP, most studies project decreases in precipitation under climate change (Argüeso et al., 2012b; Kilsby et al., 2007). However, there are other primary variables related to drought conditions through the occurrence of land-atmosphere feedbacks such as temperature, evapotranspiration, or soil moisture (Quesada et al., 2012; Seneviratne et al., 2010). This fact is particularly important over the IP, considered as a transitional region between dry and wet climates. In this context, the study of García-Valdecasas Ojeda et al. (2020b), through the analysis of projected changes of land-surface and atmospheric variables involved in the hydrologic and energy balance, has revealed that the IP is likely to experience a soil dryness by the end of the 21$^{st}$ century, more apparent in the southern IP, and stronger under the emission scenario RCP8.5.

Additionally, note that for the study of drought phenomenon, the complex and nonlinear nature of land-atmosphere interactions in the IP (García-Valdecasas Ojeda et al., 2020b) could be addressed through the advantageous simplicity of drought indices (Manning et al., 2018). The multivariate nature of drought along with the importance of incorporating temperature in drought analysis (AghaKouchak et al., 2014; Seneviratne et al., 2012; Teuling et al., 2013) can be boarded by the inclusion of potential evapotranspiration (PET) in drought indices such as the Standardized Precipitation Evapotranspiration Index (SPEI; Vicente-Serrano et al., 2010). Contrariwise to other indices such as the Standardized Precipitation Index (SPI; McKee et al., 1993), the SPEI seems to be more accurate for detecting droughts in the context of global warming (Vicente-Serrano et al., 2016) having proved to be a better indicator for identifying drought in the IP (Vicente-Serrano et al., 2016). However, only Spinoni et al. (2018) have examined potential changes in future droughts over the IP through indices contemplating this requirement. In that study, 11 bias-adjusted high-resolution (0.11°) simulations from EURO-CORDEX (Jacob et al., 2014) were used for computing future drought projections in Europe according to a composite indicator combining the SPEI, the SPI, and the reconnaissance drought index (RDI). They found that under a moderate emission scenario (i.e., RCP4.5), droughts are projected to become more frequent and severe in the Mediterranean area, while the whole European region will be affected by more frequent and severe extreme droughts under the most severe emission scenario (i.e., RCP8.5), especially at the end of the XXI century. However, this study states the importance of taking into consideration the uncertainty introduced by the ensemble of simulations, especially over those regions that are characterized by a disagreement on the sign of the drought tendency between simulations. In this context, Guerreiro et al. (2017) tried to assess the threat of the occurrence of megadroughts in some regions of the IP. According to the IPCC AR5 (IPCC, 2014), a megadrought is defined as a very lengthy and pervasive drought, which usually persists a decade or more. Results from Guerreiro et al. (2017) revealed a high range of variability for 15 CMIP5 climate models to project future droughts in the main international basins in the IP (Douro, Tajo, and Guadiana), with most of them projecting extreme multi-year droughts by the end of the XXI century and some projecting small increases of drought conditions. Along with this, the skill of those CMIP5 climate models in representing historical drought for these basins was very variable, with some of them not showing enough persistence of dry conditions and others simulating droughts that are too long and too severe. Thus, the assessment of climate change impacts on future droughts in the IP and the investigation of their uncertainty are still challenges for drought studies in the future (Spinoni et al., 2018).

Therefore, taking into account all the previously commented considerations, this work aims to characterize future drought conditions over the IP using two drought indicators, the SPEI and the SPI. These two indices only differ in that, instead of precipitation data, the SPEI uses a simple climatic water balance (i.e., precipitation *minus* potential evapotranspiration). Therefore, the comparison between them allows directly exploring the effect of evapotranspiration on drought projections, a poorly explored aspect until now in this area. This study builds on a previous one (García-Valdecasas Ojeda et al., 2017), which assessed the added value of the WRF model to generate high-resolution climate simulations for characterizing drought conditions over the IP. The findings presented in that work provided valuable information about the validation of using WRF to further studies on drought projections. Moreover, WRF was adjusted with a specific configuration scheme for the complex orographic region of the IP, endowing this work with a valuable point of view because the previously mentioned studies did not use high-resolution projections particularly configured for our study region. To do this, WRF outputs, driven by two global climate models (GCMs) from CMIP5, have been used to compute drought indices for the near (2021-2050) and far (2071-2100) future, both under two emission scenarios, the RCP4.5 and RCP8.5 (Riahi et al., 2011, 2007; Van Vuuren et al., 2011). The projections in drought conditions thus achieved, have allowed us to analyze changes in drought characteristics (i.e., frequency, duration, and severity) from a hydrological point of view. Thus, every watershed in the IP has been classified according to its drought affectation level, which is of high interest to develop adequate adaptation and mitigation strategies to climate change.

## 2. Data and Methods

### 2.1. WRF configuration

As a continuation from García-Valdecasas Ojeda et al. (2017), the WRF model with the Advanced Research WRF dynamic core, WRF-ARW (Skamarock et al., 2008) version 3.6.1 has been used to obtain primary climate variables (i.e., precipitation and maximum and minimum temperatures). The WRF simulations were carried out using two "one-way" nested domains (Fig. 1a): the coarser domain (d01), corresponding to the EURO-CORDEX region (Jacob et al., 2014) at 0.44° of spatial resolution, and the finer domain (d02), centered over the IP at 0.088° of spatial resolution (~10 km). Both domains were configured using 41 vertical levels with the top of the atmosphere set to 10 hPa. Additionally, a set of parameterization successfully adapted to the IP was also selected (García-Valdecasas Ojeda et al., 2017).

The future simulations have been performed using two different GCMs from the Coupled Model Intercomparison Project phase 5 (CMIP5) as lateral boundary conditions, the NCAR's CCSM4 (Gent et al., 2011), and the Max Plank Institute MPI-ESM-LR (Giorgetta et al., 2013).

Among all the CMIP5 climate models with data available at an appropriate spatiotemporal resolution to run WRF, the CCSM4 and the MPI-ESM-LR were selected as they proved to be adequate to obtain high-resolution simulations over the European region (McSweeney et al., 2015). However, GCMs are commonly affected by systematic biases, so bias-corrected outputs from these climate models were finally applied to complete the regional simulations. Thus, the NCAR CESM global bias-corrected CMIP5 outputs (Monaghan et al., 2014) were used. These outputs, which were corrected following the approach proposed by Bruyère et al. (2014), are online available at https://rda.ucar.edu/datasets/ds316.1 in the format required to run the WRF model. In the same way, the outputs from the MPI-ESM-LR model were corrected following the same methodology.

To analyze future projections over the IP, the periods 2021-2050 and 2071-2100 using the emission scenarios RCP4.5 and RCP8.5 were considered, in relation to the present, using as present-day climate period from 1980 to 2014. To complete the present-day simulations, the outputs from RCP8.5 were used from 2006 to 2014. This RCP adequately describes the actual present conditions, as reported by Granier et al. (2011). These present-day simulations have proven to show an adequate performance over the IP characterizing precipitation and temperature (García-Valdecasas Ojeda, 2018; García-Valdecasas Ojeda et al., 2020a), which are the main drivers for computing the SPEI and SPI drought indices. Further details about the model setup here applied can be found in García-Valdecasas Ojeda et al. (2020b).

**2.2. Drought indices: description and analysis**

The SPI and SPEI indices have been computed in this study using the SPEI R package (Beguería and Vicente-Serrano, 2017). In this package, abnormal wetness and dryness are characterized by using normalized anomalies of precipitation for the SPI case, or a climatic water balance that considers the temperature effect through the difference between the accumulated values of precipitation and the reference evapotranspiration ($ET_0$), for the SPEI. The SPEI R code allows the formulation of both indices at different time scales. The Modified Hargreaves equation (HG-PP, Droogers and Allen, 2002), which has proven to be adequate for estimating $ET_0$ values in the IP (Vicente-Serrano et al., 2014), was selected.

Drought indices have been computed at two different time scales; the 3-month time scales, for the study of episodes related to meteorological droughts (Mishra and Singh, 2010), and the 12-month time scale, to detect hydrological droughts and their effects on river streamflow and water resources (Spinoni et al., 2015a; Vicente-Serrano, 2006). For comparative purposes, both drought indices were fitted to a log-logistic probability distribution by using the maximum-likelihood method. This guarantees that the differences between the SPI and SPEI indices will be related to the temperature effects and not to the fitted probability distribution (Vicente-Serrano et al., 2011). In this work, following other studies (Dubrovsky et al., 2009; Gu et al., 2019; Leng et al., 2015; Marcos-Garcia et al., 2017; Yao et al., 2020), we assess droughts using standardized indices in a changing climate through the parameters fitting in current conditions, taking as reference the period 1980-2014. Then, drought events have been recategorized (Table 1) following a procedure similar to Spinoni et al. (2018).

In this context, the onset of a drought event is established when dry or normal/wet conditions are followed by drought conditions (drought, severe drought, or extreme drought, namely, values of the index below -1) at least for two consecutive months. In the same way, it is considered that the event ends when the index recovers values corresponding to near normal/wet conditions (index values greater than 0). Thereby, normal or wet conditions are only taken into account to define the onset and the end of drought events. The drought events thus computed have been used to determine the temporal series of the different characteristics of droughts, i.e., duration, frequency, and severity. Duration is defined as the number of months in each drought event; severity is the absolute value of the minimum index reached in that event. And, finally, the frequency is considered as the number of events per 30 years, which coincides with the entire future periods.

Projected changes of drought have been analyzed through the Delta-Change approach (Hay et al., 2000) in terms of duration, frequency, and severity of drought events by comparison between indices, time scales, RCPs, and periods. The analysis has been performed directly comparing grid-points to prevent possible compensation errors due to the smoothing effects of averaged spatial values. Thus, the projections have been analyzed through the original rotated nested domain of 0.088° (~10 km) of spatial resolution, avoiding possible errors due to interpolation methods.

Finally, with the purpose of analyzing the impact of climate change in terms of water resources, projected changes in the different drought characteristics have been analyzed through a hybrid classification. This procedure, which is similar to that from PaiMazumder and Done (2014), facilitates the interpretation of the results, allowing us to provide valuable information for policymakers. To do this, changes in frequency, duration, and severity have been spatially aggregated using the mean values for the main river basins of the IP. Here, 12 different river basins have been considered (Fig. 1b). Such basins are the results of aggregating other smaller watersheds in some cases, which are: North Atlantic (composed by the Galician Coast, Western Cantabrian, and Eastern Cantabrian watersheds), Miño-Sil (Miño-Sil, Cávado, Ave, and Leça), Duero, Ebro, Northeastern Basins, Portugal Basins (Vouga, Mondego, Lis, and Ribeiras do Oeste), Tajo, Southeastern Basins (Júcar and Segura), Guadiana (Guadiana, Sado, Mira, and Ribeiras do Algarve), Guadalquivir (Guadalquivir, Tinto, Odiel, Piedras, Guadalete, and Barbate), Southern Basins, and finally the Balearic Islands watersheds.

## 3. Results

### 3.1. Drought characteristics for current simulations

Current simulations for SPEI and SPI indices computed at 3- and 12-month time scales have been calculated from the outputs of the WRF simulations driven by the CCSM4 and MIP-ESM-LR GCMs (hereinafter named WRFCCSM4 and WRFMPI). Fig. 2 shows the results of drought frequency, duration, and severity for the SPEI and SPI indices at 3-month time scale, for the period 1980-2014. In general, the results for this period showed a number of events between 13 and 25. Drought events were more frequent according to the SPI in both the WRFCCSM and WRFMPI simulations, meanwhile, the duration of such events (mean durations between 4 and 7 months in most of the IP) was longer for the SPEI. In any case, the results from both simulations and for both indices showed a broad common behavior in terms of drought conditions with changes in location and surface extent. Regarding severity, the values ranged from 1.2 to 1.6 in most of the IP for both simulations and indices.

For events computed at 12-month time scale (Fig. 3), current simulations generally displayed fewer events, which were longer than at the 3-month time scale (values of between 4 and 10 events and between 9 and 21 months for frequency and duration, respectively). Again, for both simulations, the SPI showed more events, which resulted shorter as well. In reference to the severity, the events were more moderate, with a greater number of grid points reaching lower values than at shorter time scale, at least for the WRFCCSM. Nevertheless, the magnitude of such values was in a similar range of values.

**3.2. Projected changes in drought parameters for near future**

Changes in the frequency, duration, and severity of drought events, for the near future (2021-2050) relative to the current period (1980-2014), for the SPI and SPEI indices computed at 3- and 12-month time scales, from WRFCCSM4 and WRFMPI simulations under the RCPs 4.5 and 8.5, are presented in this section.

Fig. 4 shows this analysis for RCP4.5 at 3-month time scale. All WRF simulations driven by the intermediate GHG emissions scenario projected both increases and decreases in the number of events, for the entire near future, in a range from about -10 to 10 events (Fig. 4, left column). The SPEI in WRFCCSM indicated a widespread increase in the frequency except over certain scattered regions located mainly over the Duero and Tajo Basins (Central IP). This same simulation, but using the SPI, revealed more decreases in the number of events than for the SPEI in most of the watersheds. Exceptions were the North Atlantic and Balearic Islands Basins, where the number of events is similar to that from SPEI with increases of around 7 events/30 years. In the same way, the WRFMPI simulation projected both increases and decreases, with larger areas presenting a reduction in the number of events by using both the SPEI and the SPI for most of the IP.

Changes in the mean duration of such events (Fig. 4, second column) showed moderate increases, overall, in all WRF simulations (values ranging from about -3 to 4 months). The lengthening of the average duration of the drought events proved slightly greater for the SPEI for both WRFCCSM and WRFMPI. In general, the changes in severity (Fig. 4, third column) were positive in all simulations (values up to 0.6), although many areas presented almost an absence of changes with respect to this parameter. The most notable increases again appeared in the SPEI for WRFCCSM, with values of around 0.1-0.6, practically through the entire IP, with the highest values being located mainly in the Duero Basin and over southern watersheds (i.e., Guadalquivir and Southern Basins). Such increases in values indicate that the drought events in many regions of IP become severe or extreme since, in general, the severity values in the present were around 1.4 (see Fig. 2), so increases of 0.1-0.6 signify mean values of over 1.5 for the near future.

For RCP8.5, at 3-month time scale, Fig. 5 reveals similar spatial frequency patterns to those shown under RCP4.5, with changes in the same range of magnitude as well. For WRFMPI, using the SPEI, striking increases (changes > 6 months) in terms of duration were found over the southwest of the Guadiana Basin and in a large part of the Guadalquivir Basin, both located in the southern third of the IP. Here, the increase of severity was also substantial with respect to RCP4.5, reaching values up to 0.6 relative to the present period. Again, the SPI, for the two simulations, presented more moderate values of change than the SPEI.

At 12-month time scale (Fig. 6 and 7 for RCP4.5 and RCP8.5, respectively), changes in the near future presented by the two indices showed a broader common spatial behavior than those at 3-month time scale, but with a greater magnitude for the SPEI. Under RCP4.5 (Fig. 6), WRFCCSM showed drought events more frequent for the near future in relation to the present in many parts of the IP (increases of up to 7 events/30 years in a large part of the IP). The Tajo Basin appeared to be the most affected by the increase in the frequency of drought events, reflected especially by the SPEI. By contrast, the WRFMPI simulation presented a generalized decline in frequency to around 5 events for a large part of the IP. Exceptions of such behavior were found in the Southeastern Basins and in a part of the northwestern IP (i.e., North Atlantic and Miño-Sil Basins), where increases of around 5 events were reached.

On average, the results also showed that drought events are likely to be longer in many parts of the IP (changes of more than 12 months), being this particularly marked for the SPEI. In this way, both indices presented major increases in the mean duration, especially over the Duero and Guadalquivir Basins. In terms of severity, in general, changes for SPEI were generally stronger than those found for SPI. Here, the WRFCCSM projects decreases as well as increases (values of around -0.6 and 0.8), with the growing severity occurring mainly in the eastern areas and northern Portugal. By contrast, the WRFMPI under this scenario appeared to show more extended increases, covering a large area of the IP. As an exception here, a part of North Atlantic watersheds showed less severity.

In terms of frequency, the patterns of change for the RCP8.5 in the near future (Fig. 7) are very similar to those found in the intermediate emission pathway forcing, although the number of events appeared to be slightly moderate. For duration and severity, however, and as occurred at 3-month time scale, the changes were also more moderate for the WRFCCSM, and substantially more marked for the WRFMPI simulation. In the latter, a large area over the south of the peninsula (i.e., the Guadalquivir, Guadiana, and Southern Basins) presented quite long events, lasting more than 12 and 24 months on average for the SPI and SPEI, respectively. These long values also appeared in certain areas over the Southeastern Basin as well as the watersheds of the Tajo and over watersheds located in the northeastern part of the IP (i.e., Ebro, Northeastern, and Balearic Islands). According to the WRFMPI simulation, the severity is likely to increase throughout nearly the entire IP (values up to 0.8) except in certain regions over the Northeastern Basins as well as in some parts of the Miño-Sil and North Atlantic Basins (both in the northwest), for both SPEI and SPI indices.

**3.3. Projected changes in drought parameters for far future**

Fig. 8 shows the projected changes for the period 2071-2100 for the indices at 3 months under RCP4.5. For this period, drought conditions are expected to be greater in magnitude than for the near future, in general. Thus, the CCSM4-driven simulation showed greater frequency, reaching values above 15 events per period throughout basins in the southern IP (i.e., Guadalquivir and Guadiana), as well as in the North Atlantic watershed, in the northernmost part of the peninsula. The Duero Basin also appeared more affected by drier conditions than in the near future. However, other watersheds such as the North Atlantic, Southeastern, Ebro, and Portugal Basins presented a great surface area with changes as great as in the near future. Here, for the SPI, again, less pronounced changes were found than for the SPEI, in general. For the WRFMPI nevertheless, an increase in the number of drought events appeared in watersheds in the north (changes around 10 events/period), and a decline in the number of events was found in southern and southeastern IP watersheds (reductions by around 7 events/period) in general.

The results also revealed an increase in the mean duration, showing more affected areas with longer events for the SPEI, and especially for the simulation driven by the MPI-ESM-LR (values above 10 months in some regions). For this parameter, the WRFCCSM projected the longest events in eastern watersheds (i.e., Ebro, Southeastern, and Balearic Islands Basins) and at some points in the Guadalquivir and Southern Basins. Whereas, the WRFMPI indicated increases particularly marked in the basins of the southern half of the IP, such as the Guadalquivir, Guadiana, and Southeastern Basins. The severity was also projected to increase reaching values of around 0.6 in practically the entire IP for the SPEI in both the WRFCCSM and the WRFMPI.

Under RCP8.5, the results at 3 months in the far future (Fig. 9) revealed a lower number of events in several regions of the IP from the analysis of the SPEI for the WRFCCSM simulation (values of change between -10 and 15 events/period in practically all the IP). Meanwhile, a prevalence of large areas with increases was found for the SPI from this same simulation. By contrast, the WRFMPI projected changes similar to those from RCP4.5 for both indices, and with approximately the same range of values as well. However, in terms of duration, marked changes were found, particularly for the SPEI. In the latter, the southern half of the IP underwent marked increases of more than 12 months. Although the increase in severity was also quite pronounced throughout the IP in both simulations and for the two indices, the strongest severities (increases up to 0.8) were projected by the SPEI from the WRFCCSM simulation.

As occurred at 3-months, drought events at 12-month time scale in the far future under RCP4.5 (Fig. 10) presented change patterns similar to those projected for the near future. In terms of changes in frequency, the values remained similar to those simulated for the near future in the WRFCCSM (values between -10 and 7 events per period) for both indices. However, in this period, the number of events for the entire period was slightly lower particularly in the Ebro Basin, in the northwestern part of the peninsula. For the MPI-ESM-LR-driven simulation (with changes in the same range), the increase in the number of drought events was limited fundamentally to certain regions over the northern half of the IP (i.e., North Atlantic, Ebro, Miño-Sil, and Portugal Basins) for the SPEI, and also in certain areas along the Duero watershed for the SPI. By contrast, the rest of the IP showed a lower number of drought events than in the present period in general.

On the other hand, substantial increases in the mean duration were also found for this period (increases of 72 months or higher). The longest events were located mainly in the northwestern IP (Ebro and Northeastern Basins) for the WRFCCSM and for both indices. Additionally, for the SPEI, other regions also suffer these long events throughout the IP (Balearic Islands, Guadiana, Guadalquivir, Tajo, Duero, Southern, and Southeastern Basins). WRFMPI showed similar spatial patterns but with more pronounced increases over the entire IP. Thus, for the SPEI, certain parts mainly in the east of the Guadalquivir Basin as well as in the Southeastern and Ebro watersheds (in the eastern IP), presented an increase in the mean duration of the drought events of 96 months (i.e., 8 years), or more.

In terms of severity changes, the simulations driven by either the CCSM4 or the MPI-ESM-LR under RCP4.5 (Fig. 10) projected different drought patterns. The WRFCCSM indicated more moderate increases (around 0.6), which do not affect the entire IP. In this case, the most affected areas appeared in the northwest (i.e., Miño-Sil and Portugal Basins), northeast (i.e., the Ebro, Balearic Islands, and Northeastern Basins) as well as in certain parts of the central and southern IP (e.g., Duero, Guadalquivir, Southern, and Southeastern Basins). Meanwhile, for WRFMPI, the greater severity spread over practically the entire IP, with increases of up to 0.8 or more.

For RCP8.5 in the far future, the changes at 12-month time scale (Fig. 11) were extremely strong. The frequency was substantially reduced (changes from about -10 and 7 events for the overall period), which is likely associated with the extraordinary increase in the mean duration. For the WRFCCSM simulation, results from the SPEI showed a generalized decrease of as many as 5 events in most of the IP, the total number of events, therefore, being reduced to 1 or 2 events over the entire period in many cases (see Fig. 3). For the SPI, decreases were also shown in general except for scattered regions (e.g., increases of around 5 events/30 years related to the present period in the northwest of the Ebro Basin). In the simulations driven by MPI-ESM-LR, although the overall trend is also to reduce the number of events, this resulted less marked, showing broader areas with increases, located in the northwest of IP, for both indices.

Substantial changes in terms of duration were found, especially for the SPEI. For this index, both the WRFCCSM and the WRFMPI showed regions with increases of more than 96 months (8 years). The most pronounced changes were projected by the WRFCCSM simulation, in which most of the IP presented drought events longer than 10 years (120 months), even reaching values of 180 or higher in many of the watersheds. The SPI, however, presented more moderate changes in duration, although in any case, these were substantial as well. Therefore, these results suggest that by the end of the century and under a scenario where the emission of GHGs is especially high, the potential risk to suffer megadroughts is very high, or the dramatic changes in precipitation and temperature could lead to greater aridity in the IP. Again, this evidences the importance of taking into account the temperature to analyze potential changes in the aridity conditions. In terms of severity, all simulations showed a generalized increase throughout the IP, with values being above 0.8, which rose for the SPEI WRFCCSM simulation.

### 3.4. Hybrid classification in drought event characteristics

Finally, a hybrid classification for the three parameters of drought events has been performed (Figs. 12 and 13). To this end, the frequency, duration, and severity previously detailed have been spatially averaged for each river basin within the IP (see Fig. 1b) and, thus, different categories have been established based on whether such characteristics increase or decrease in relation to the present values.

For the near future (Fig. 12), some uncertainties appeared in the sign of the change in drought characteristics as was indicated by the results found through the use of different driving data. In this context, the results from the WRFMPI showed a signal more robust, with similar patterns of change for both time scales in most of the watersheds of the IP. Here, the river basin least affected appeared to be the North Atlantic basins. By contrast, the WRFCCSM simulations suggested a more different trend depending on the RCP, drought index, and time scale, although the results from the SPEI indicated drier conditions in general, as was previously explained at both time scales. In any case, all the results showed an increase of at least one drought characteristic, the duration increase being the most prevalent.

For the far future (Fig. 13), the sign of the change was clearer and more robust, as is reflected by the results from the different simulations, indices, and time scales, with the North Atlantic basins, in any case, being the least affected. However, although the changes were different in magnitude between scenarios, the sign of the change was similar for both RCPs in most of the watersheds. For this period, the most prevalent characteristics were the increases in the duration and severity.

**4. Discussion**

This work constitutes a continuation of a previous study, in which the added value of the WRF model to simulate drought conditions in the IP was evidenced (García-Valdecasas Ojeda et al., 2017). Now, based on that proved ability, this study aims to explore high-resolution drought projections for a near (2021-2050) and a far (2071-2100) future under different RCPs. For this end, the WRF outputs, using two bias-adjusted simulations from the CCSM4 and MIP-ESM-LR GCMs as lateral boundary conditions, which include climate projections for the RCP4.5 and RCP8.5 (García-Valdecasas Ojeda et al., 2020a, 2020b), have been used. As in García-Valdecasas et al. (2017), drought events have been defined according to the SPEI and SPI indices computed for 3- and 12-month accumulation periods.

Present-day simulations revealed a similar range of values for drought events characteristics than those from observations (Figs. 1S and 2S in the supplementary material). However, drought indices from simulations indicated longer events than the observed ones in certain regions and depending on the index and the GCM-driven WRF simulation. Subsequently, simulated drought events must be less frequent as well. These features have been previously noted in other works (Burke et al., 2006; Guerreiro et al., 2017). Also, note that some of the discrepancies between the observed and simulated drought characteristics can result from the different periods used to compute the drought indices (1980-2014 for drought events computed from simulations vs. 1980-2010 for the observed ones) and due to the fact that observational gridded products here used are also affected by inherent errors, which can be occasionally large (Gómez-Navarro et al., 2012).

Projections of drought conditions here found agree with the projections for temperature and precipitation (García-Valdecasas et al., 2020a) in the near future. That is, in both cases, WRFCCSM revealed spatial similar changes between the RCPs and even slightly more severe for the intermediate RCP forcing; while WRFMPI pointed out moderate changes under RCP4.5 which become substantial for RCP8.5.

These same WRF simulations indicated substantial changes by the end of the century in primary climate variables such as temperature, precipitation, surface evapotranspiration, and soil moisture, particularly under RCP8.5 (García-Valdecasas Ojeda et al., 2020b), so that marked differences in future drought conditions in relation to the present are also expected for the IP climate. In this regard, note that using standardized drought indices to assess changes in drought phenomena with pronounced changes in dryness conditions could be inaccurate to suitably quantify the projected changes, as pointed out by Guerreiro et al. (2017). However, certain valuable information can be considered by adopting a categorized new classification for drought conditions. So, in this context, we find results similar to those reported by Guerreiro et al. (2017), in general terms. These authors, using the Drought Severity Index (DSI) at a 12-month time scale, found a marked increase in the length of drought, corresponding to multi-year drought events, for the Duero, Tajo, and Guadiana watersheds. Also, similar results, overall, are found to that reported by Spinoni et al. (2018), which used the entire period 1981-2100 as baseline for fitting the drought indices, finding that droughts are projected to become increasingly more severe in the IP, especially after 2070 and under RCP8.5. However, while they established more frequent drought in the IP, in this work longer droughts but less frequent are stated in the future. This partial discrepancy could be due to the different calibration periods considered to estimate drought indices, which is currently a key issue in drought assessment to better understand regional drought characteristics and the associated temporal changes, particularly under climate change scenarios (Um et al., 2017). Note that frequency and duration for a given period are inversely correlated so longer events become less frequent, anyway the increase in either duration or frequency could indicate an increase in drought events.

Moderate changes in drought events have been found in the near future, particularly in terms of duration, with minor differences between scenarios. By contrast, by the end of the 21st century, drier conditions are expected, with noteworthy differences in relation to the present. In this period, the differences between RCPs are also evident. In fact, while the results from RCP4.5 suggest a downturn in the upward trends, notable increases are found for RCP8.5, indicating that drought conditions are likely to become more common by the end of the century. In relation to the spatial patterns of the changes, similar results are found in the simulations driven by both GCMs for the two periods and scenarios. This fact suggests a relatively robust response in terms of drought events. In this context, the magnitude of such changes is determined by the period and emission scenario. These results partially agree with those of Stagge et al. (2015) and Spinoni et al. (2018), who found an increase in the drying conditions by the end of the century over the IP by computing drought indices from an ensemble of EURO-CORDEX projections. In Stagge et al. (2015), the authors pointed out a progression in dry conditions under RCP8.5, while for RCP4.5 the drought indices reached maximum values for the period 2041-2070.

Concerning the comparison between indices, the results clearly corroborate the importance of taking into account the effect of the temperature to assess the impact of climate change for the future. Thus, projections in drought events using the SPI show more moderate changes than those from the SPEI, especially for the far future. This is because an index based solely on precipitation cannot explain the full magnitude or spatial extent of drying reflected by the SPEI (Cook et al., 2014). In fact, the expected rises in temperature lead to greater moisture demand by the atmosphere and, consequently, increased evapotranspiration, which could result in even more severe impacts than precipitation deficits in a warmer world (Ault et al., 2016). In the far future, for the higher emission scenario, simulations showed a substantial rise in temperatures as well as a reduction in precipitation, indicating a strong joint effect. This has been pointed out by many authors (Ault et al., 2016; Burke et al., 2006; Dai, 2013; García-Valdecasas Ojeda et al., 2020a, 2020b; Marcos-Garcia et al., 2017). In particular, for the IP, dryness conditions are mainly driven by reductions in precipitation, but consequences are seriously intensified by higher temperatures (García-Valdecasas Ojeda et al., 2020a, 2020b).

The results from drought indices computed for the end of the century, and especially for the longest time scale (12-months) and for the SPEI, suggest a serious risk of megadrought events. In fact, the drought indices evaluated at the 12-month time scale provide additional information on the general trend over time since the accumulated values of either precipitation or water availability for each new month have less impact on the total amount, the response of the index being more slowly (McKee et al., 1993). Therefore, the longer duration here means the stabilization in drier conditions. In this sense, drought events from the SPEI at 12-month are extremely long in the far future (more than 15 years in many cases), suggesting that the IP could likely undergo a megadrought, in accordance with the definition provided by Ault et al. (2016). That study defined a megadrought as an event in which Palmer Drought Severity Index (PDSI) values fall below -0.5 standard deviations for a period of at least 35 years. Although our study period is somewhat shorter than 35 years, the results found here from the 12-month SPEI, which is and drought index analogous to the PDSI (Vicente-Serrano et al., 2010), could suggest that the IP will follow trends towards this kind of drought. These results could also indicate a change in the aridity conditions, namely, the values that are below normal conditions in the present (rare events or extremes) could become normal in the future. This agrees in general terms with the study of Gao and Giorgi (2008), who examined projected changes in arid climate regimes by computing three different measures of aridity using high-resolution projections over the Mediterranean region. They found that this region will likely undergo a notable increase in dry and arid land under increased GHG concentrations, particularly in regions such as the IP. In this context, our results could also indicate that PET effects could intensify and expand the drying northwards from the Mediterranean.

Our findings, based on two GCMs under two RCPs, pointed out a clear trend in the future drought conditions in the IP, at least in the sign, as shown in the results from the hybrid classification. However, note that projections are affected by certain limitations and uncertainties, especially for time horizons of several decades. These are mainly associated with the GCM behavior to reproduce climate conditions over a region and the different socioeconomic scenarios that may happen in the future (Hawkins and Sutton, 2011).

**5. Conclusions**

Globally the results of this study have shown that a generalized increase in drought conditions for the IP is expected. However, at a high spatial resolution, substantial differences in drought characteristics have been found for the future, depending on the studied Basin. The main findings of this study are as follows:

- The IP is very likely to undergo longer and more severe drought episodes in the future. Substantial changes in drought characteristics have been projected by both indices and time scales. Such changes are probably to be particularly strong by the end of the century under the higher emissions scenario (RCP8.5) when greater duration and severity of drought events in relation to the present have appeared in most of the IP. The latter is even more striking in terms of hydrological droughts (i.e., indices computed at 12-month time scale). However, the intensification of drought conditions remains more moderate in the near future. In this period, the results have revealed a certain degree of uncertainty between the GCM-driven simulations for some areas, while the difference between RCPs has been less marked. These findings suggest slow GHGs induced climate change effects for the near future.

- There are highlight differences in evaluating drought events using an index based solely on precipitation data (SPI) and another one that takes into account the effect of the temperature rise (SPEI). Projected drought conditions by using the SPEI have shown more severe increases in drought events than those from SPI by the end of the century and, especially, for the high-emission scenario.

- The IP might suffer extremely long drought periods. Large parts of the IP has shown increases in mean duration in relation to present conditions of around 10 years (or more), for the period 2071-2100 under RCP8.5. This indicates that the drought indices values that are below normal conditions in the present (rare events or extremes) could become normal in the future result of an increase in aridity or the occurrence of megadrought events.

- The assessment of future droughts from a river basins point of view can help for the development of adequate mitigation and adaptation strategies for water management under climate change in the IP. Thus, the study of the changes by using a hybrid classification has shown more severe drought conditions in the future, especially by the end of the XXI century and over the Mediterranean Iberian river basins. Here, an agreement regarding the sign of the changes between the different GCM-driven simulations has suggested a robust climate change signal.

- Despite the limited number of simulations analyzed (using just one RCM driven by two GCMs), these results could serve as a starting point for estimating the impacts of future drought events, and consequently, for the development of adequate mitigation and adaptation strategies for water management under climate change in the IP.

**Acknowledgments**

This work was financed by the FEDER / Junta de Andalucía - Ministry of Economy and Knowledge / Project [B-RNM-336-UGR18], and by the Spanish Ministry of Economy, Industry and Competitiveness, with additional support from the European Community Funds (FEDER) [CGL2013-48539-R and CGL2017-89836-R]. We thank the ALHAMBRA supercomputer infrastructure (https://alhambra.ugr.es) for providing us with computer resources. The first author is supported at present by OGS and CINECA under HPC-TRES program award number 2020-02.We thank the anonymous reviewers for their valuable comments that helped to improve this work.

**Figures**

**Fig. 1**. a) Domain of the study. The 0.44º EURO-CORDEX domain (d01) as coarser domain and the inner 0.088º domain (d02) centred over the Iberian Peninsula. b) Hydrographic basins of the Iberian Peninsula.

**Fig. 2.** Drought frequency (F, left), duration (D, middle), and severity (S, right) for current period (1980-2014) for the SPEI and the SPI indices computed at 3-month time scale. Duration and severity were obtained from values averaged for the entire period whereas the frequency is the number of events for the entire period.

**Fig. 3.** As Fig. 2, but for indices computed at 12-month time scale.

**Fig. 4.** Changes in the frequency (ΔF, left), duration (ΔD, middle), and severity (ΔS, right) of drought events for the near future (2021-2050) relative to the current period (1980-2014) and under RCP4.5. Drought events are based on indices computed at 3-month time scale.

**Fig. 5**. As Fig. 4, but for simulations driven by the GCMs under RCP8.5.

**Fig. 6.** Changes projected for the near future in the drought frequency (events/30 years), duration (months), and severity for the indices computed at the 12-month time scale under RCP4.5.

**Fig. 7.** As Fig. 6, but for simulations driven by the two GCMs using RCP8.5 forcing.

**Fig. 8.** Changes in the drought frequency (ΔF), duration (ΔD), and severity (ΔS) for indices computed at 3-month time scale for the far future (2071-2100) related to the present period (1980-2014) and under RCP4.5.

**Fig. 9.** As Fig. 8, but for the simulations driven under RCP8.5.

**Fig. 10.** Projected changes in drought frequency, duration, and severity from indices computed at the 12-month time scale in the far future and for RCP4.5.

**Fig. 11.** As Fig. 10, but for the simulations driven under the RCP8.5 scenario.

**Fig. 12.** Hybrid classification based on projected changes in severity, frequency and duration (ΔS, ΔF and ΔD, respectively) of droughts according to the SPEI and the SPI at 3- and 12-month time scales, for the near future.

**Fig. 13.** As Fig. 12 but for the far future.

Table1Table 1 Drought categories for the study of drought events.

| Drought index value | Drought Category | Conditions |
|---|---|---|
| index $\leq -2$ | -2 | extreme drought |
| $-2 <$ index $\leq -1.5$ | -1.5 | severe drought |
| $-1.5 <$ index $\leq -1$ | -1 | drought |
| $-1 <$ index $\leq 0$ | -0.5 | near normal |
| $0 \leq$ index | 1 | wet |

**(a)** Figure 1

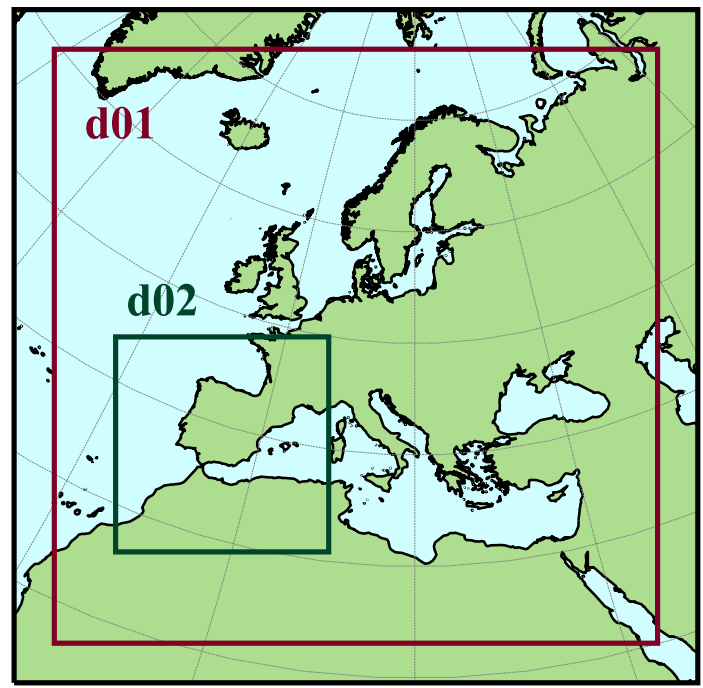
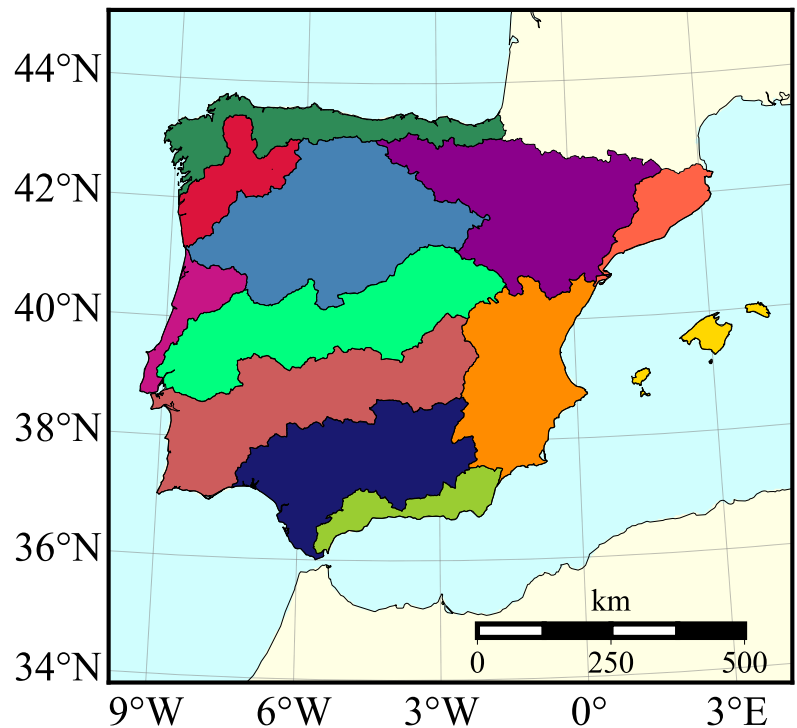
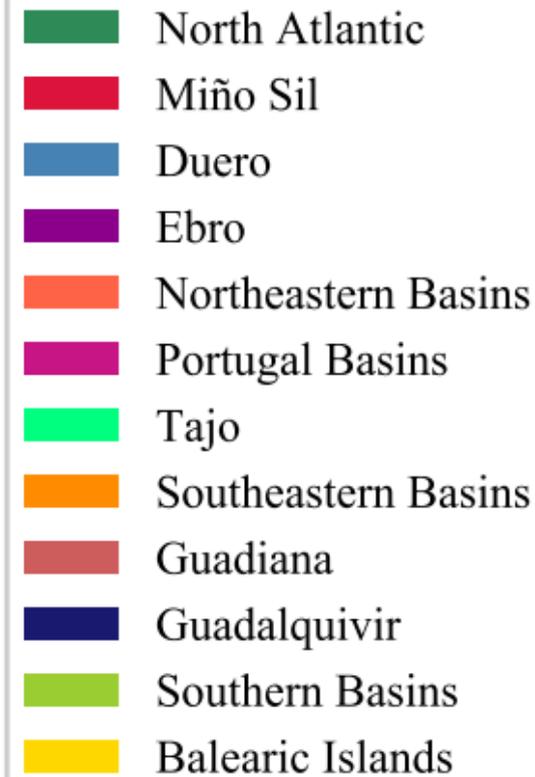

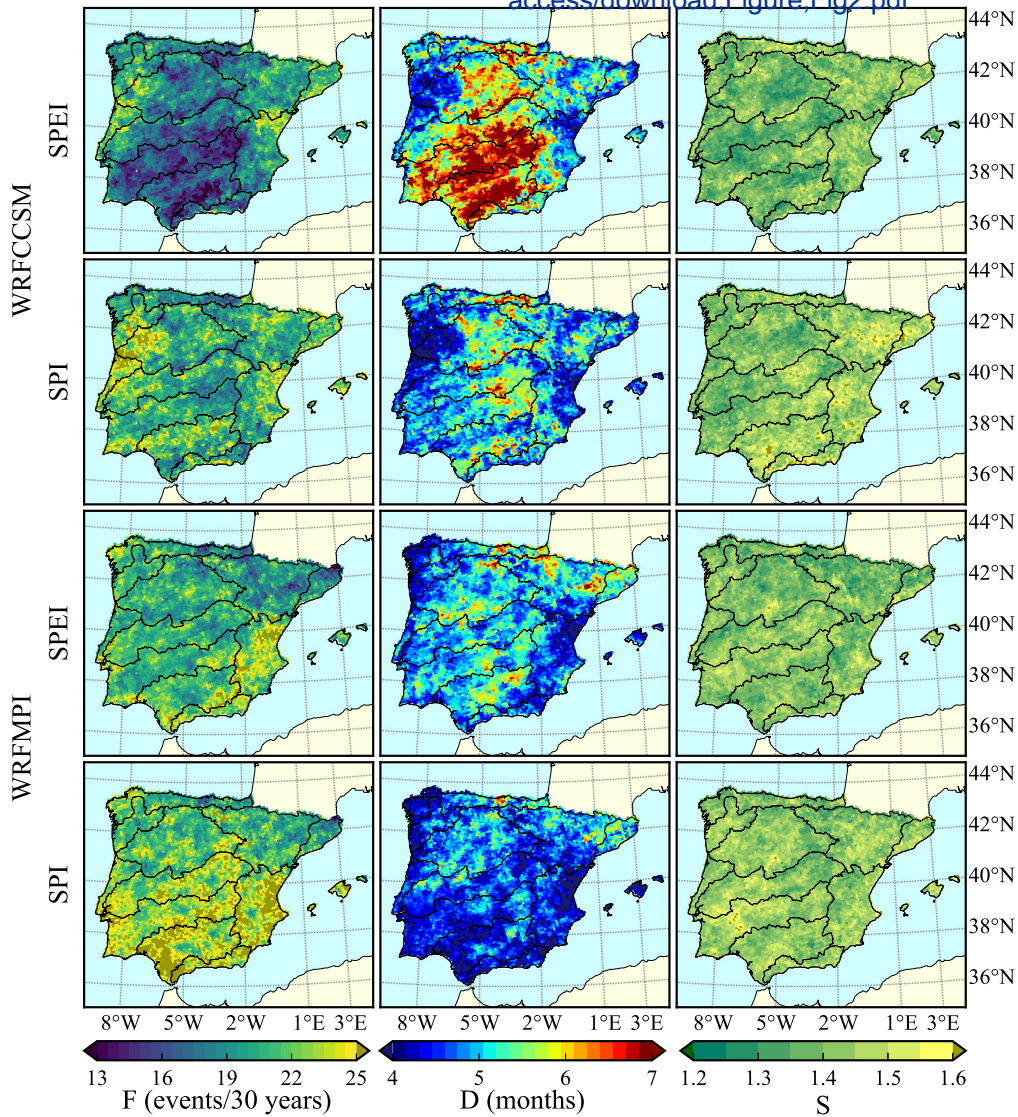

Figure 2

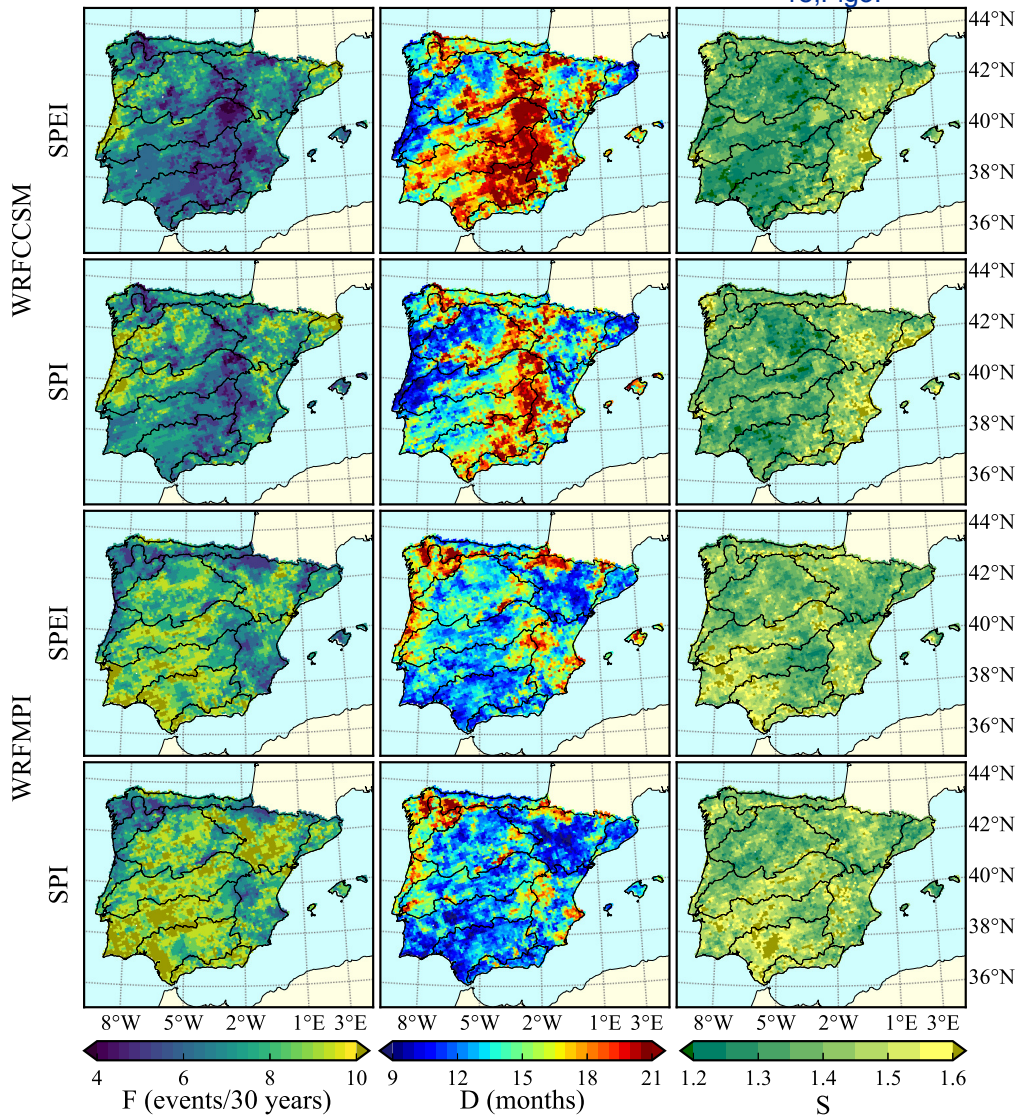

Figure 3.

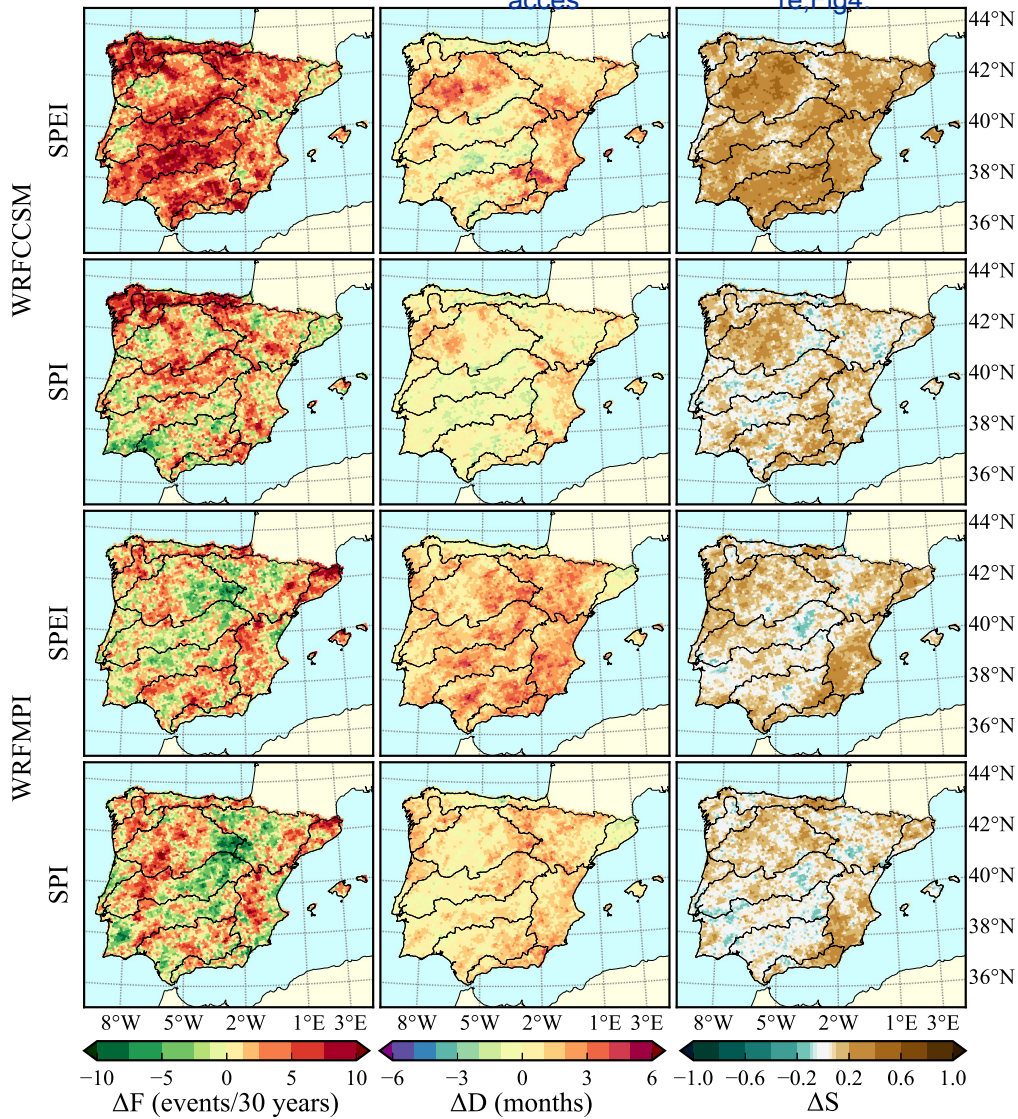

Figure 4

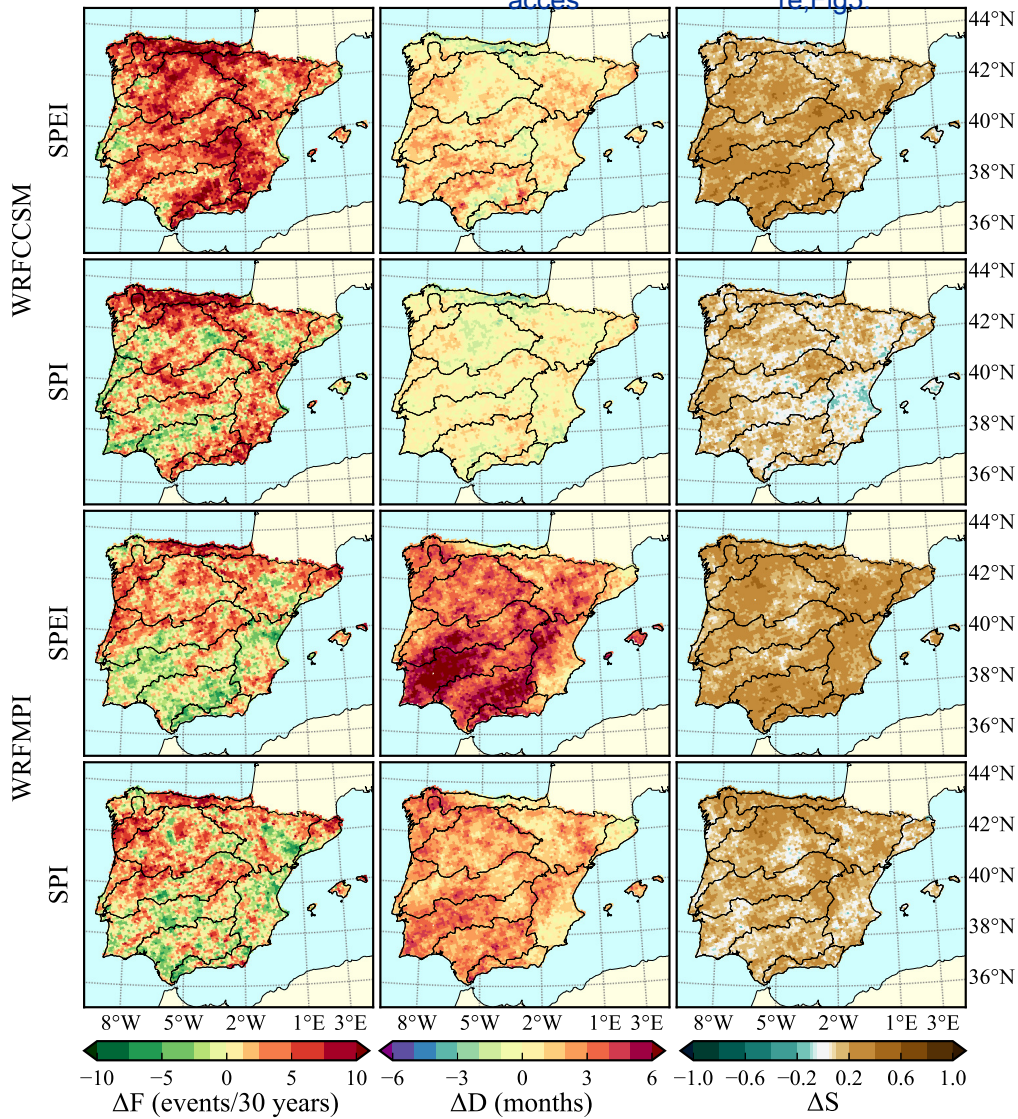

Figure 5

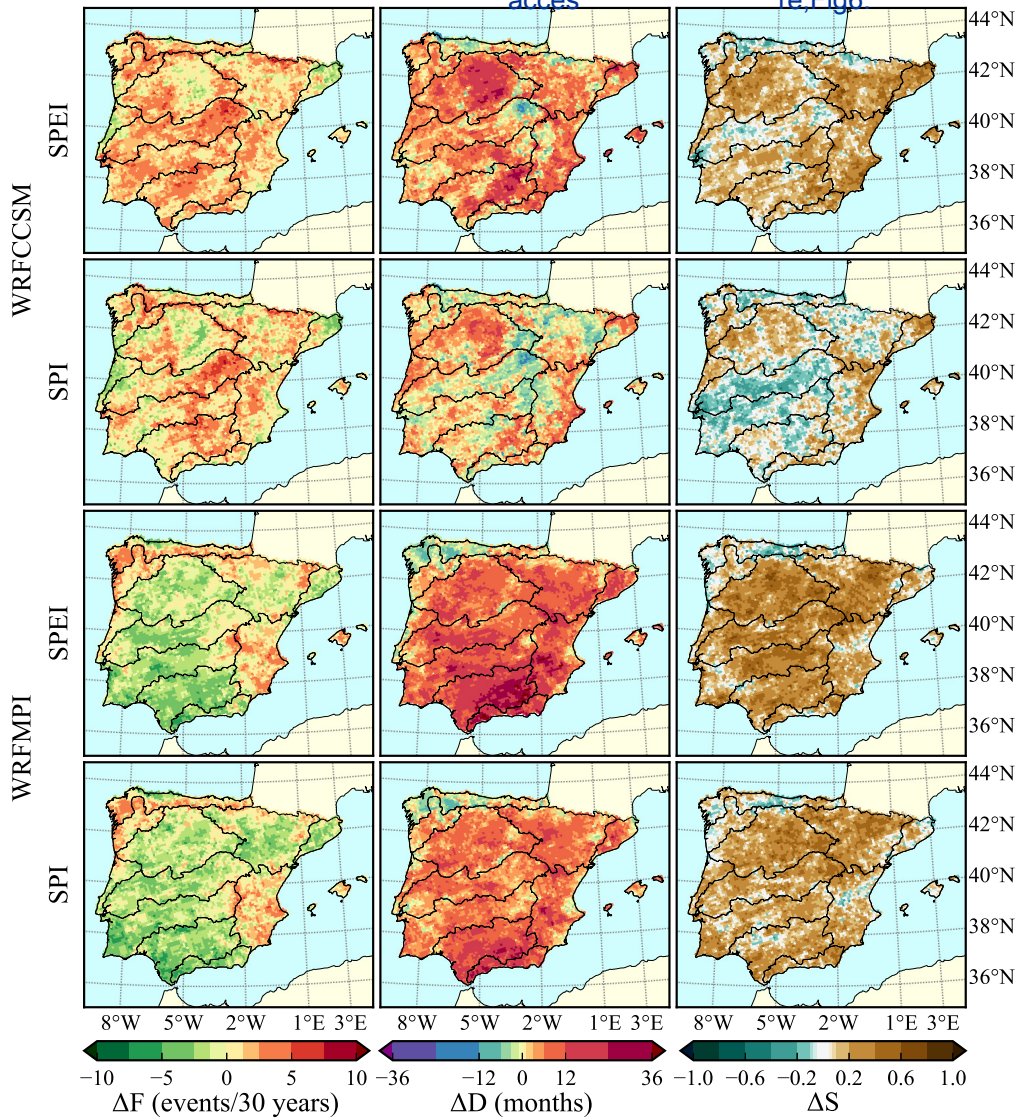

Figure 6

Figure 7

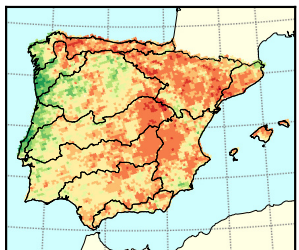 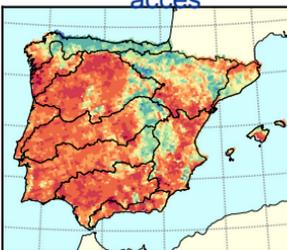 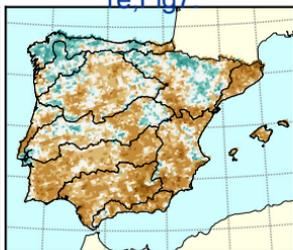
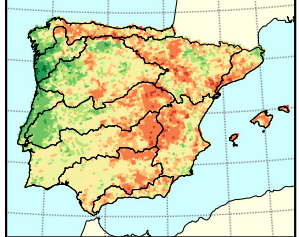 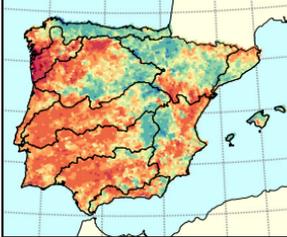 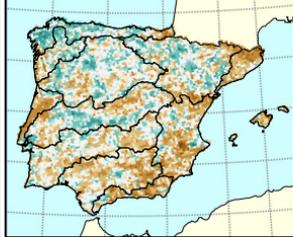
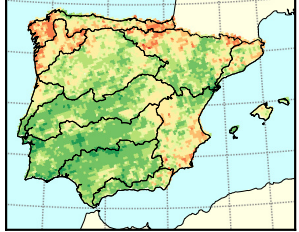 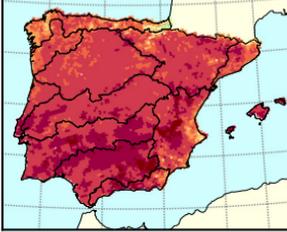 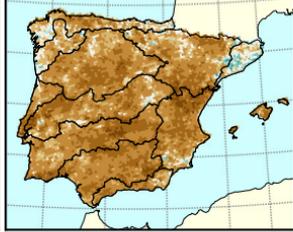
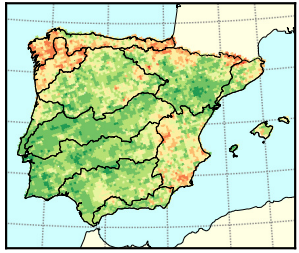 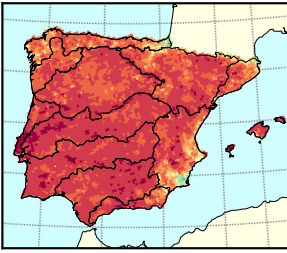 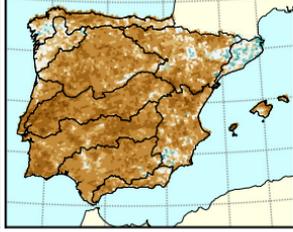
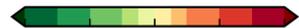 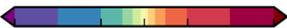 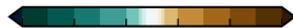

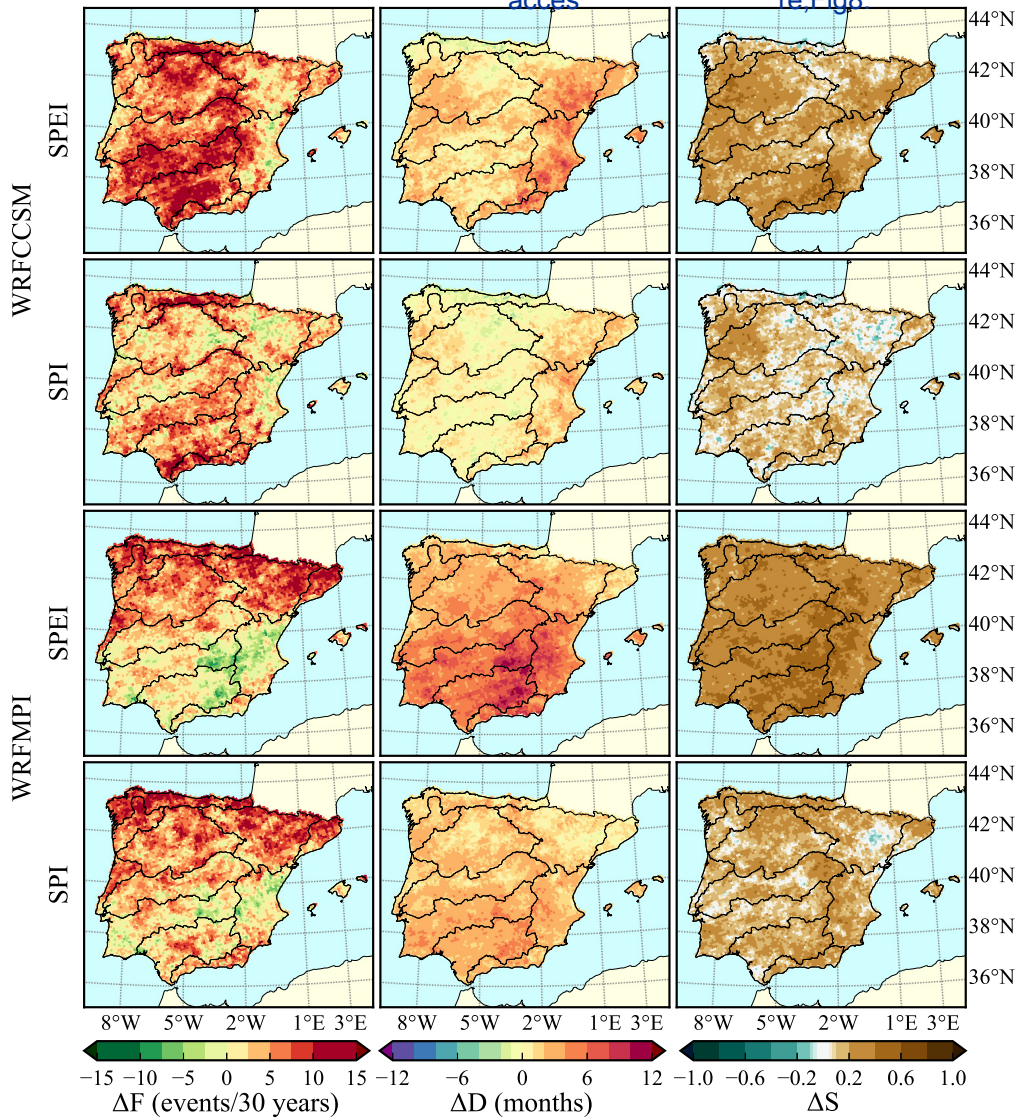

Figure 8

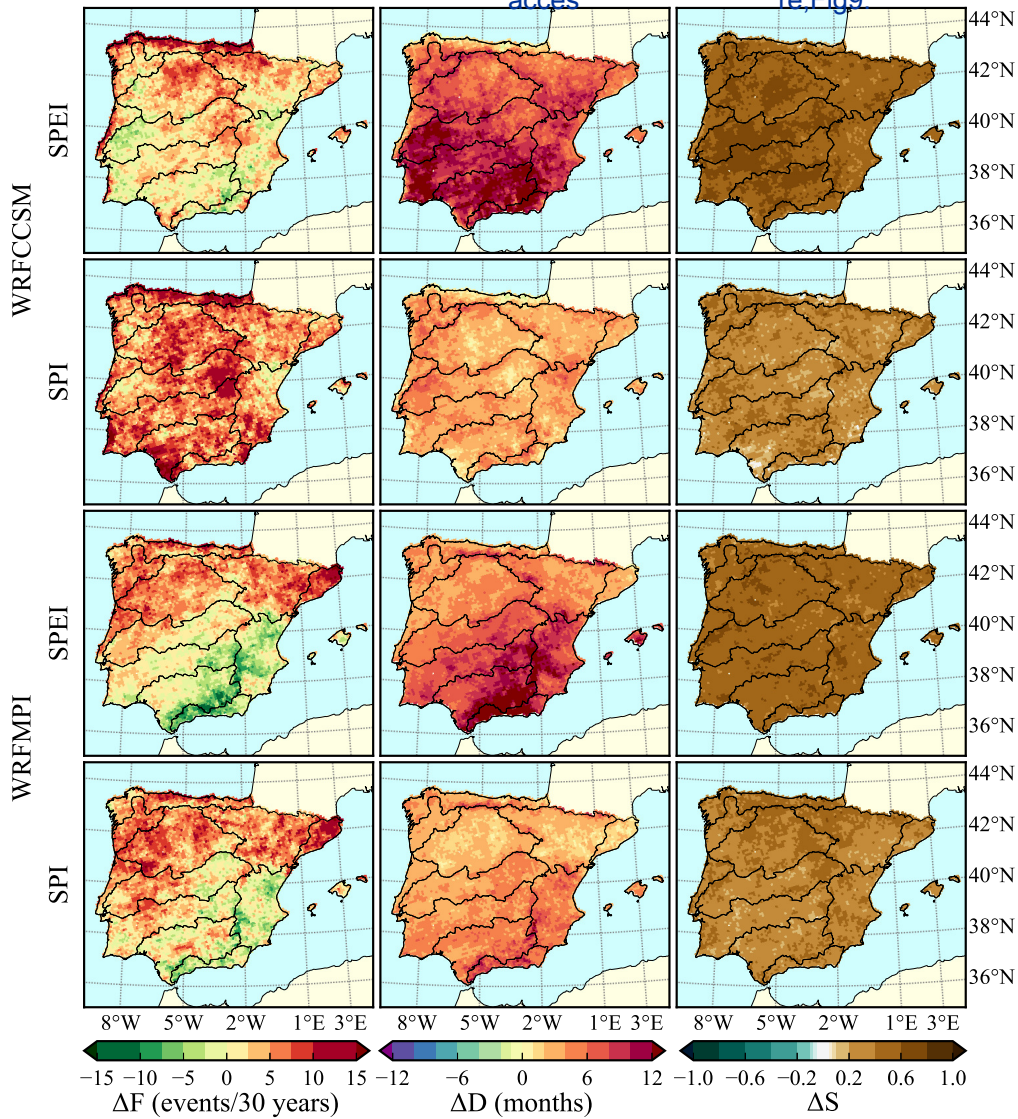

Figure 9

Figure 10

| | Frequency | Duration | Severity |
|---|---|---|---|
| WRFCCSM / SPEI | 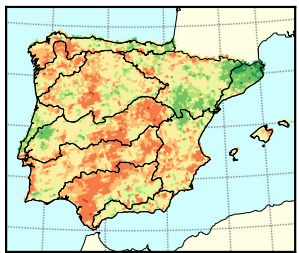 | 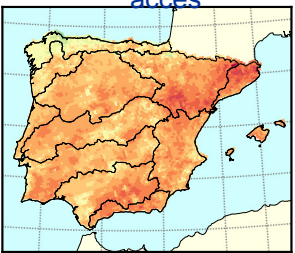 | 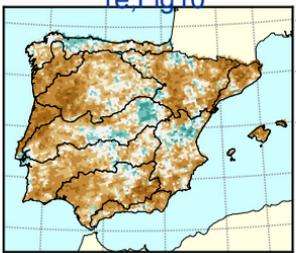 |
| WRFCCSM / SPI | 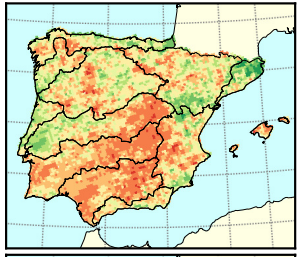 | 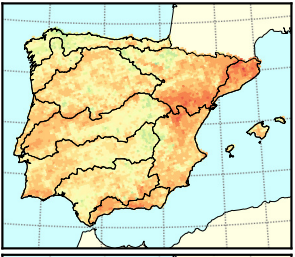 | 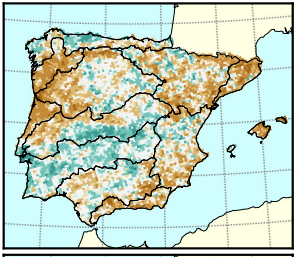 |
| WRFMPI / SPEI | 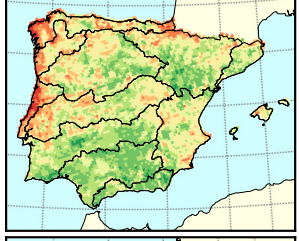 | 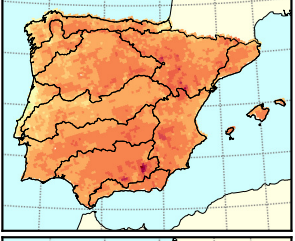 | 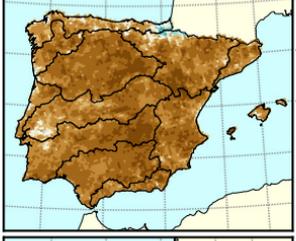 |
| WRFMPI / SPI | 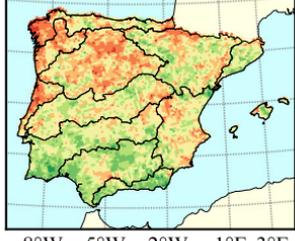 | 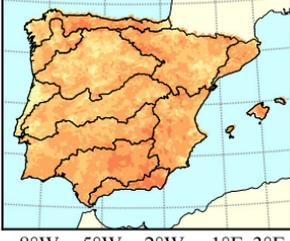 | 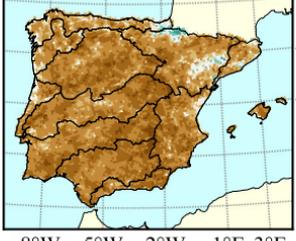 |

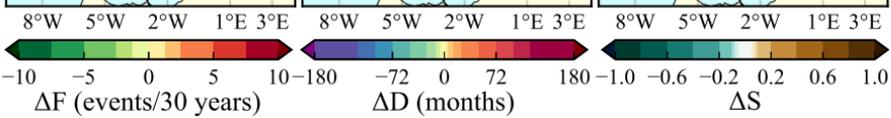

Figure 11

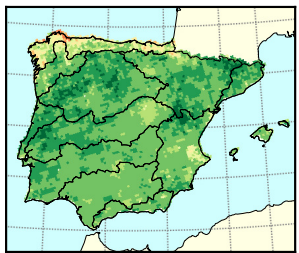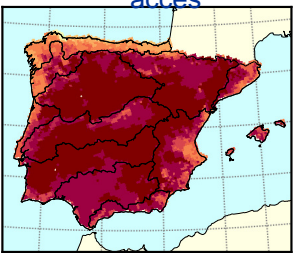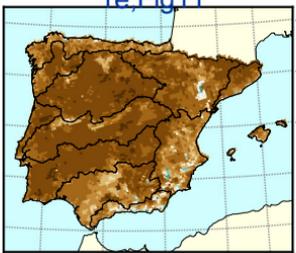
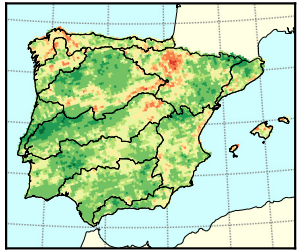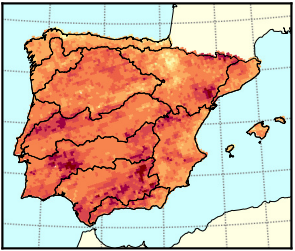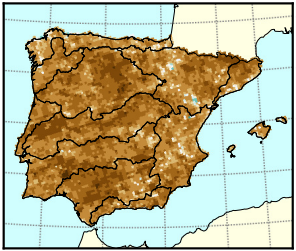
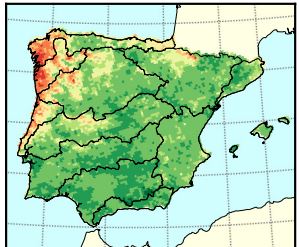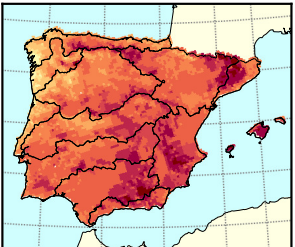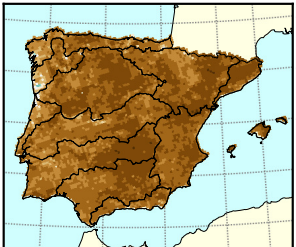
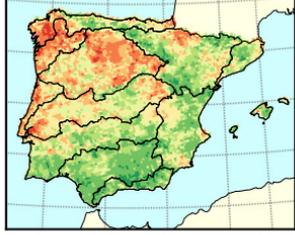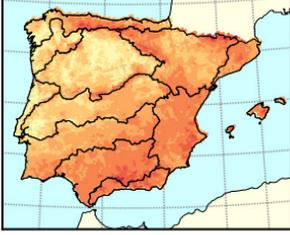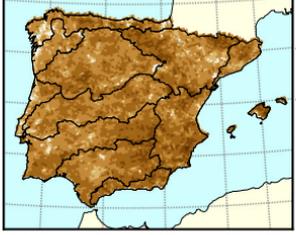
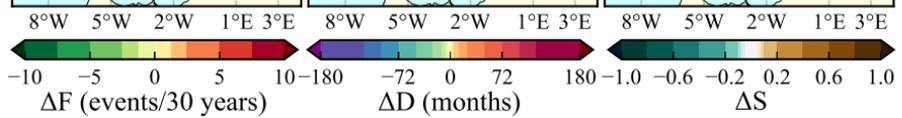

Figure 12

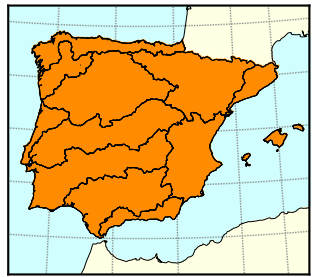 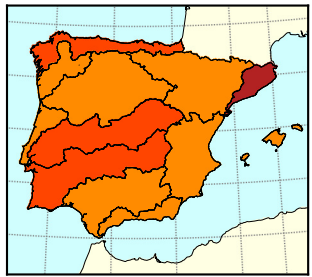 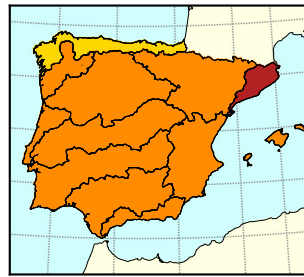 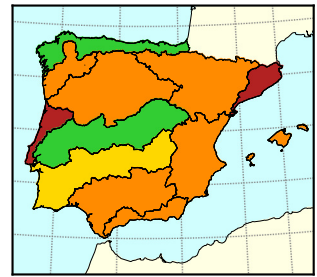
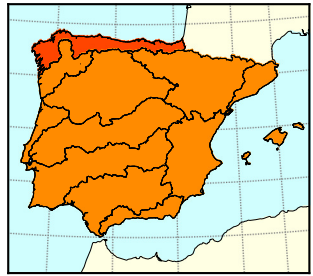 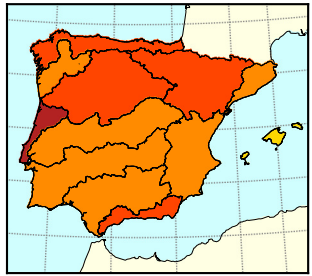 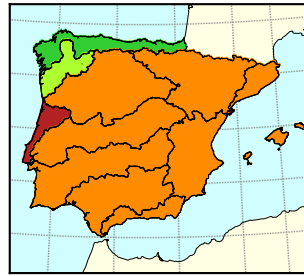 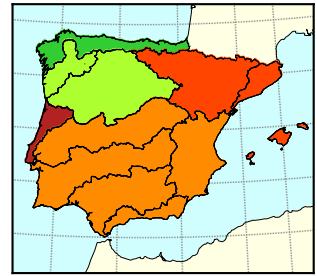
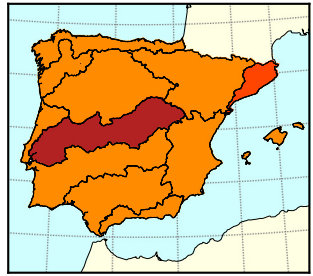 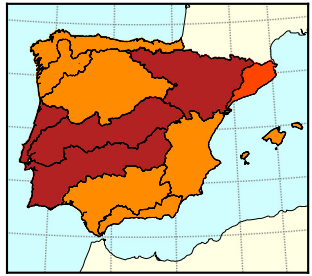 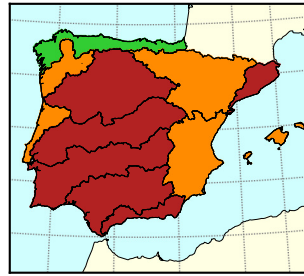 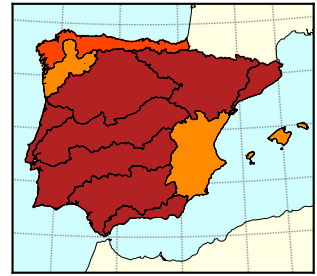
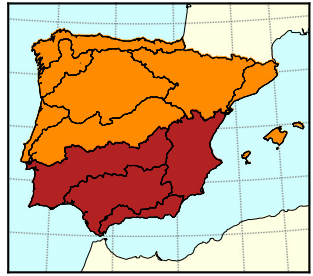 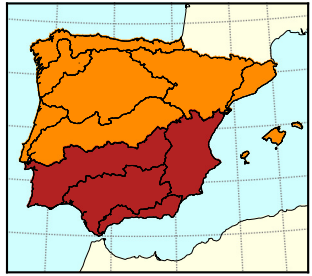 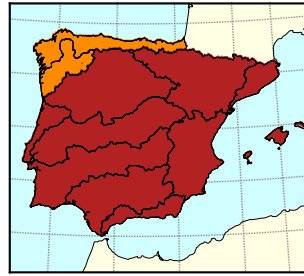 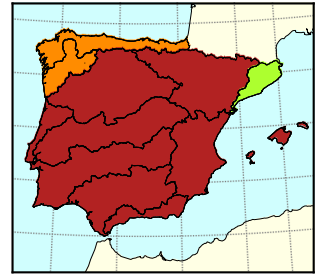

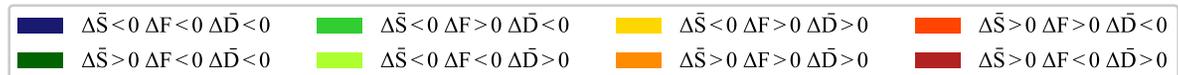

Figure 13 

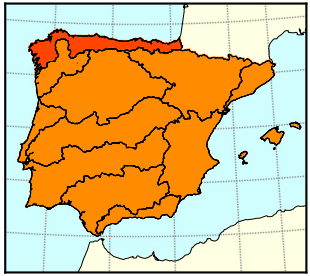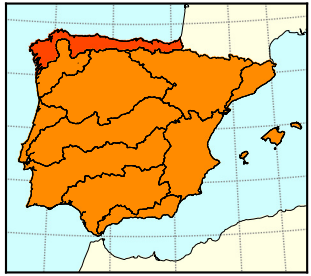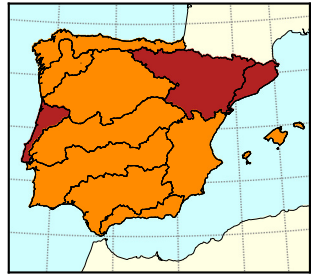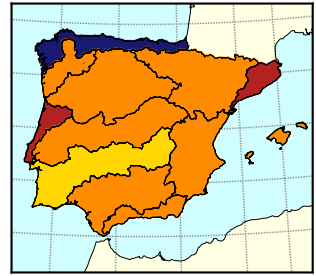
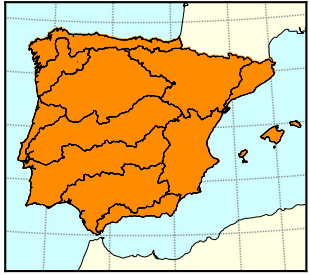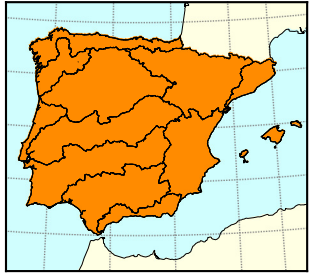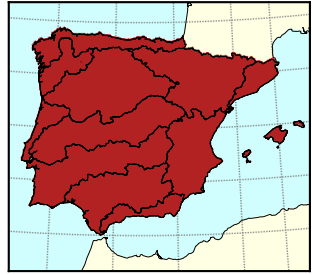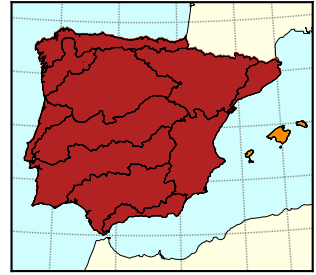
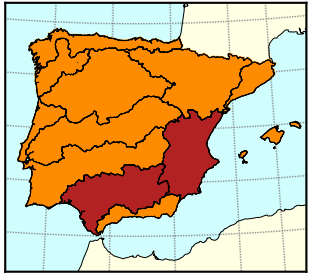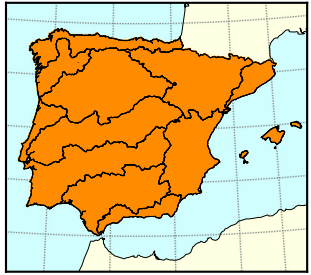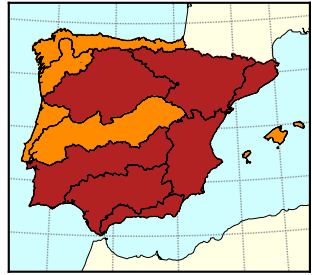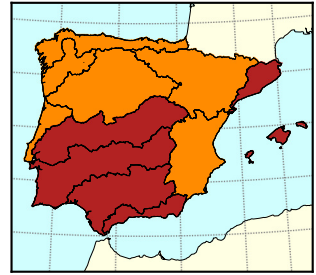
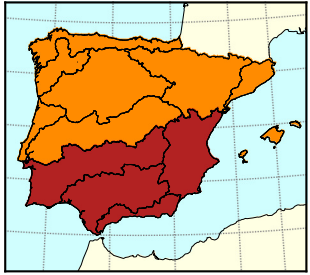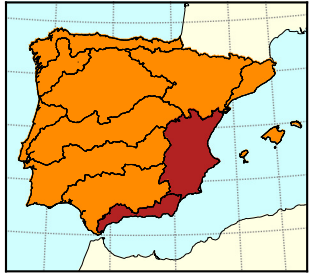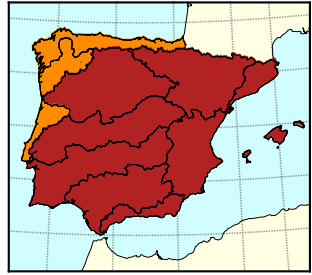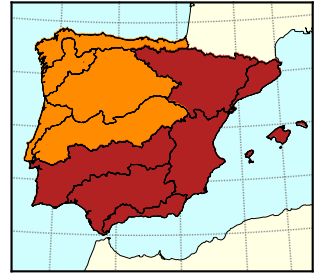
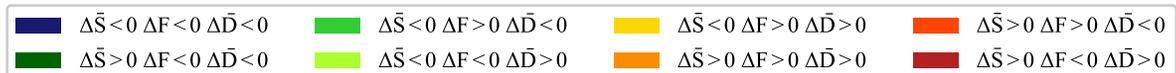